\documentclass[reprint,superscriptaddress,aps,pra,nofootinbib]{revtex4-2}
\usepackage[utf8]{inputenc}
\usepackage[english]{babel}
\usepackage[T1]{fontenc}
\usepackage[colorlinks=true, allcolors=blue]{hyperref}

\usepackage{amsmath}
\usepackage{dsfont}
\usepackage{physics}
\usepackage{xcolor}
\usepackage{graphicx}
\usepackage{float}
\usepackage{enumitem}
\usepackage{amsfonts}
\usepackage{comment}
\usepackage{mathtools}

\DeclarePairedDelimiter{\ceil}{\lceil}{\rceil}

\begin{document}

\title{Reflection-Based Adiabatic State Preparation}

\author{Jessica Lemieux}
\thanks{Corresponding author: \href{mailto:jessica.lemieux@1qbit.com}{jessica.lemieux@1qbit.com} }
\affiliation{D\'epartement de Physique, Universit\'e de Sherbrooke, Sherbrooke, QC, Canada}
\affiliation{Institut Quantique, Universit\'e de Sherbrooke, Sherbrooke, QC, Canada}
\affiliation{1QB Information Technologies (1QBit), Sherbrooke, QC, Canada}

\author{Artur Scherer}
\affiliation{1QBit, Waterloo, ON, Canada}

\author{Pooya Ronagh}
\affiliation{1QBit, Vancouver, BC, Canada}
\affiliation{Institute for Quantum Computing, University of Waterloo, Waterloo, ON, Canada}
\affiliation{Department of Physics \& Astronomy, University of Waterloo, Waterloo, ON, Canada}
\affiliation{Perimeter Institute for Theoretical Physics, Waterloo, ON, Canada}

\begin{abstract}
  We propose a circuit-model quantum algorithm for eigenpath traversal that is based on a 
  combination of concepts from Grover's search and adiabatic quantum computation. 
  Our algorithm deploys a sequence of reflections determined from eigenspaces 
  of instantaneous Hamiltonians defined along an adiabatic schedule in order to prepare 
  a ground state of a target problem Hamiltonian. We provide numerical evidence suggesting that, 
  for combinatorial search problems, our algorithm can find a solution faster, on average, than 
  Grover's search. We demonstrate our findings by applying both algorithms to solving the \mbox{NP-hard} 
  \textsc{MAX-2SAT} problem.
\end{abstract}

\maketitle

\section{Introduction}
Grover’s search~\cite{grover1996fast} is a famous quantum algorithm for unstructured search that 
offers a provable quadratic speed-up in comparison to exhaustive classical search. Variants of the 
algorithm have been studied since its invention 25 years ago~\cite{aaronson2020quantum,boyer1998tight,
brassard2002quantum,durr1996quantum}. Grover’s algorithm was proven to be optimal in finding a 
marked element in an unstructured problem~\cite{zalka1999grover}. However, 
many industrially relevant real-world computational problems have 
additional structures, and exploiting them could result in potentially
more-efficient heuristics that may outperform Grover’s search. 
Combinatorial and discrete optimization problems are 
examples of such structured problems. In our research, we consider the Boolean satisfiability
problem as our working example. In particular, we use the NP-hard \textsc{MAX-2SAT} problem for 
our numerical studies. 

We present a reflection-based adiabatic algorithm (RBA) based on a combination of concepts from Grover’s 
search and discrete adiabatic state preparation~\mbox{\cite{aharonov2003adiabatic,farhi2001quantum,lemieux2021resource}}.
Our algorithm resembles eigenpath traversal navigated by the quantum Zeno effect~\cite{boixo2009eigenpath,boixo2010fast}, 
but we replace projective measurements along the eigenpath with reflections to guide the evolution. In Sec.~\ref{subsec:ET}, we provide a discussion on the relationship between our research and previous work. 
The implementation of the RBA could follow any eigenpath traversal, such as along the instantaneous ground states of an adiabatic evolution. The algorithm’s sequence of reflections can be interpreted as slowly changing the marked state(s) when conducting a search using Grover's algorithm.

To study the potential of the RBA, we benchmark its performance against that of Grover’s algorithm for 
small instances of the \mbox{\textsc{MAX-2SAT}} problem. We believe that the RBA can also be successfully 
applied to solving optimization problems with non-classical objective functions. Examples of such problems include preparing the ground state of a generic
Hamiltonian for fermionic systems, a problem known to be \mbox{QMA-hard~\cite{kempe2006complexity,kitaev2002classical,oliveira2005complexity}}.

\section{Algorithm}
The RBA uses a sequence of reflections
$\left(R_1,R_2,...,R_L\right)$ defined by the ground states of
a sequence of respective Hamiltonians. Let us denote the $k$-th ground state in this sequence by $\ket{G_k}$,
where $k\in\{1,\dots,L\}$. The algorithm consists of the following main
steps:
\begin{enumerate}
  \item Prepare the initial state $\ket{\text{\texttt{Init}}}$.
  \item For $k=1, \dots, L$, apply the reflection
  \mbox{$R_k := \mathds{1} - 2 \ketbra{G_k} $}.
  \item Perform a measurement in the eigenbasis of the target problem
  Hamiltonian.
\end{enumerate}
Depending on the type of problem, steps 1--3 may need to be repeated.
We expand on the details of these steps in what follows.

\subsection{Intermediate Hamiltonians}
To define the ``good'' subspaces through which the reflections
are applied, we introduce intermediate Hamiltonians along an adiabatic path,
although our approach allows more-generic paths.
For simplicity, we use a linear interpolation between the
starting Hamiltonian $H_0$ and a problem Hamiltonian $H_1$:
\begin{align}
  H_w := (1-w)H_0 + wH_1.
\end{align}
A discretization in time corresponds to a sequence of weights $\left(w_1,w_2,...,w_L \right)$
 that specify the instantaneous
Hamiltonians $\left(H_{w_1},H_{w_2},...,H_{w_L}\right)$.
The ground states of these instantaneous Hamiltonians, respectively,
characterize the corresponding good subspaces. The choice of the weights $w_k$ could also be optimized classically by 
minimizing the expected energy of the resulting states, defining a hybrid 
quantum algorithm. These kinds of hybrid quantum--classical algorithms are 
often studied in the context of NISQ and variational quantum algorithms~\cite{farhi2014qaoa,peruzzo2014variational}. The classical component of these 
hybrid schemes is a nontrivial optimization problem that is expected to scale 
poorly. For example, the barren plateau~\cite{holmes2021connecting,mcclean2018barren}
is the phenomenon of gradient norms decreasing exponentially fast, causing the 
need for exponentially many gradient estimations to be made with regard to 
problem size. Therefore, in Sec.~\ref{subsec:numStudy} we present the results of our study on the decay rate of the gradient norms for our hybrid scheme.

\subsection{Reflections}
The implementation of reflections requires information about the
eigenstates of the intermediate Hamiltonians, which we do not have.
However, this information can be coherently acquired by performing
quantum phase estimation~(QPE) in the basis of the respective intermediate
Hamiltonian, each time a reflection is deployed. 
In order to do this
coherently without collapsing the state, the QPE step is followed by
an energy value comparison with a classical threshold $E^*$.
This energy threshold could either be stored in an additional 
quantum register or be ``hard-coded'' in the arithmetic needed
for the comparison. The result of the comparison is stored in a flag qubit. In other words, instead of measuring the energy register,
which would collapse the superposition state, we obtain a flag qubit
entangled with the eigenstates, such that the qubit is 
in the state $\ket{1}$ when the eigenstates correspond to an energy
below the threshold $E^*$, and in the state $\ket{0}$ otherwise. 
Effectively, this output corresponds to marking the state
around which the reflection is performed.
Recall that QPE is an 
algorithm for estimating the phases associated with the eigenvalues 
of a unitary operator. In our analysis, the unitaries are given by 
$U_k=e^{iH_{w_k}}$, which have the same eigenstates as $H_{w_k}$.
The phases that correspond to the eigenvalues of $H_{w_k}$ are denoted by $E^j_{w_k}$
for every $j$-th eigenvalue.
Thus, to ensure that QPE differentiates between all
eigenstates, the Hamiltonians $H_0$ and $H_1$ must be normalized such that
$0 \leq E_{w_k}^j < 2\pi$ for all $k$ and $j$. 
Note that this approach requires us to have lower and upper bounds for the energy spectrum. 

In what follows, we denote the $j$-th eigenstate of $H_{w_k}$ by $\ket{\psi_{w_k}^j}$.
The reflection $R_k$, \mbox{for $k=1, \dots, L$,} can then be implemented as follows.
\begin{enumerate}
  \item Deploy QPE with $U_k:=e^{iH_{w_k}}$ and an additional register of size $M$ 
  used to hold the estimated energy value:
  \begin{equation}
  \text{QPE}\ket{0}^M\ket{\Psi} = \sum_{j=0}^{2^n-1}
  \braket{\psi_{w_k}^j}{\Psi}\ket{E_{w_k}^j}\ket{\psi_{w_k}^j}.
  \end{equation}
  Here, $\ket{\Psi}$ is an arbitrary state and $\ket{E_{w_k}^j}$ represents the
  energy value corresponding to the eigenstate $\ket{\psi_{w_k}^j}$ of $H_{w_k}$.
  \item Apply an energy value comparison with a classical threshold $E^*$, which requires arithmetic. Append a single-qubit ancilla initialized in the computational
  state $|0\rangle$. If and only if $E_{w_k}^j < E^*$, flip the ancilla.
  If $E^*$ is chosen such that $E_{w_k}^0 < E^*\le E_{w_k}^1$ (assuming the ordering\footnote{In the event of the ground states being degenerate, the threshold is chosen to be between the ground state energy and the next energy level.}
  $E_{w_k}^0 < E_{w_k}^1 \leq E_{w_k}^2 \leq \dots $), this results in
  the entangled state
  \begin{align}
  \braket{G_k}{\Psi}\ket{E_{w_k}^0}\ket{G_k}\ket{1} &\nonumber \\
   +\sum_{j=1}^{2^n-1}\braket{\psi_{w_k}^j}{\Psi}&\ket{E_{w_k}^j}\ket{\psi_{w_k}^j}\ket{0},
  \end{align}
  where $\ket{G_k}=\ket{\psi_{w_k}^0}$.
  \item Apply the Pauli-$Z$ gate to the flag ancilla, which results in
  a negative phase for the first term: \\
  \begin{align}
  -\braket{G_k}{\Psi}\ket{E_{w_k}^0}\ket{G_k}\ket{1} &\nonumber \\+\sum_{j=1}^{2^n-1}\braket{\psi_{w_k}^j}{\Psi}&\ket{E_{w_k}^j}\ket{\psi_{w_k}^j}\ket{0}.
  \end{align}
  \item Uncompute the registers for the energy comparison and QPE to yield the following:
  \begin{equation}
   -\braket{G_k}{\Psi}\ket{G_k}
  +\sum_{j=1}^{2^n-1}\braket{\psi_{w_k}^j}{\Psi}\ket{\psi_{w_k}^j}.
  \end{equation}
\end{enumerate}
\begin{figure*}
  \centering
  \includegraphics[width=0.8\linewidth]{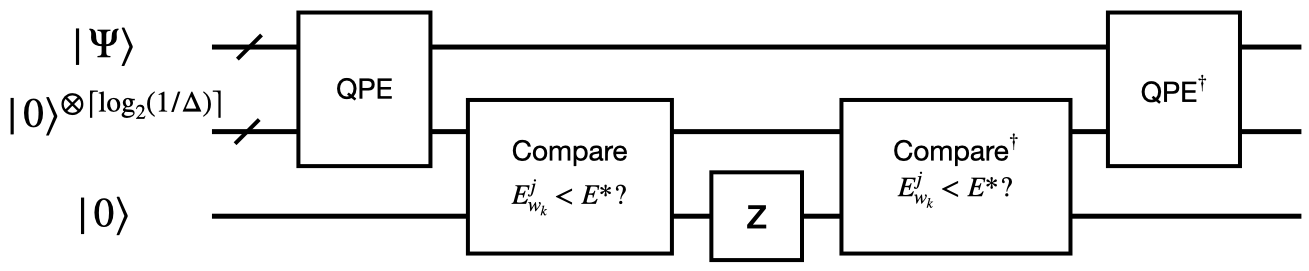}
  \caption{Quantum circuit for implementing a reflection}
  \label{fig:RefCircuit}
\end{figure*}
A quantum circuit diagram for these steps is shown
in Fig.~\ref{fig:RefCircuit}. This figure illustrates that
the implementation cost of a reflection is not significantly different
from that of a projective measurement.
A projective measurement would terminate after QPE deployment (step 1)
by measuring the register holding the energy value (cf.~\cite{boixo2010fast,lemieux2021resource}).
To ensure that QPE
is able to differentiate between the ground state and the first excited state,
we need to estimate the phases $E^j_{w_k}$ with a precision high enough to distinguish between $E_{w_k}^0$ and $E_{w_k}^1$. 
In other words, the number of qubits required to hold the energy values 
obtained from QPE must scale as $M \in \mathcal{O}\left(\ceil{\log_2\left(1/\Delta_k\right)}\right)$, where \mbox{$\Delta_k := E^1_{w_k}-E^0_{w_k}$} is the $k$-th energy gap.
As QPE scales exponentially with respect to $M$,
this leads to an overall scaling of
$\mathcal{O}(1/\Delta)$ in terms of a lower bound on the energy gaps, 
$\Delta\le \Delta_k $, for all \mbox{$k=0, \dots, L$.} With respect to the reflection, adding the energy comparison scales with the number of
qubits as $M \in \mathcal{O}\left(\ceil{\log_2\left(1/\Delta_k\right)}\right)$.
The uncomputation of the ancillary registers (necessary for reversible
computation) doubles the cost. 
 Thus, a projective measurement
and a reflection have roughly the same scaling in terms of query complexity, which is  $\mathcal{O}(1/\Delta)$.

There are multiple alternatives for the implementation of QPE.
Normally, this algorithm requires Hamiltonian simulation of the unitary $U_k=e^{iH_{w_k}}$. This
is typically performed using techniques based on
Lie--Trotter product formulae~\cite{berry2007efficient},
truncated Taylor series~\cite{berry2015simulating},
quantum signal processing~\cite{low2017optimal}, or
qubitization~\cite{low2019hamiltonian}.
Two difficulties are usually encountered when implementing QPE
by digitally simulating the operator $U_k=e^{iH_{w_k}}$:
the error introduced by the Hamiltonian simulation and
ambiguities in the phase. Using the framework of
qubitization~\cite{low2019hamiltonian}, these difficulties
can be eliminated by replacing $U_k$ with the qubiterate (see Refs.~\cite{berry2018improved,lemieux2021resource,poulin2018quantum}). Since the energy spectrum is changed by qubitization, using the qubiterate will affect the query complexity of the reflections. The query complexity becomes $\mathcal{O}\left(1/\arccos\left(1-\Delta\right)\right)$ instead of $\mathcal{O}(1/\Delta)$~\cite{lemieux2021resource,low2019hamiltonian}.
There are also techniques that incorporate the linear combination of unitaries or
oblivious amplitude amplification~\cite{chowdhury2018improved}.
However, in view of the actual purpose of our work, we do not include
these additional improvements. For simplicity, our analysis is based
on the original approach of implementing QPE by simulating the unitary
$U_k$.
\subsection{Connection to Grover's Search Algorithm}

Grover's search~\cite{grover1996fast}, and the related amplitude amplification algorithm~\cite{brassard2002quantum}, use two reflections: one through the subspace
spanned by the superposition state in the ``good'' subspace,
$\ket{\texttt{Succ}}$, and
another one through the subspace
spanned by the initial state, $\ket{\texttt{Init}}$:
\begin{align}
  R_1^G = \mathds{1} - 2 \ketbra{\texttt{Succ}} = e^{i \pi \ketbra{\texttt{Succ}}}\,, \\
  R_2^G = \mathds{1} - 2 \ketbra{\texttt{Init}} = e^{i \pi \ketbra{\texttt{Init}}}.
\end{align}
The algorithm   repeatedly applies $R_1$ followed by $R_2$, whose combined
effect is called the Grover iteration.
The key uncertainty is in figuring out when to stop to achieve the highest
probability of success.
If \mbox{$|\braket{\texttt{Succ}}{\texttt{Init}}| = \sin(\theta/2)$,}
the optimal number
of Grover iterations is \mbox{$n^{\text{\tiny opt}}_{\text{it}}
= \lceil\frac{\pi}{2\theta}-\frac{1}{2}\rceil$,}
in which case the algorithm outputs a solution state with a success
probability equal to $\sin^2\left[\left(n^{\text{\tiny opt}}_{\text{it}}+1/2\right)\theta\right]$.
However, we generally do not know the overlap 
$|\braket{\texttt{Succ}}{\texttt{Init}}|$. Since the probability of success
is periodic in $n_{\text{it}}$ in this scheme, exceeding
$n^{\text{\tiny opt}}_{\text{it}}$ results
in its decrease. As for discrete adiabatic state preparation~\cite{boixo2009eigenpath}, in our scheme,
increasing the number of intermediate steps will most likely
increase the probability of success.

On the other hand, Grover's algorithm can also be viewed as a special case
of the RBA, where the weights $w_k$ alternate between $0$ and $1$.
Alternatively, imagine a perfectly defined reflection that
maps the initial state to the desired target state in a single step.
It can be shown that the gap of the intermediate Hamiltonian corresponding to
such a reflection would be exactly $\sin(\theta/2)$ (similar to the proof made 
by Roland and Cerf~\cite{roland2002quantum}).
Thus, implementing this reflection using the steps described in the previous
section would lead to the same scaling, because the gap determines
the precision needed for QPE. Moreover, implementing a Trotter--Suzuki
decomposition of this reflection, with
$R_1^G$ and $R_2^G$ as the composing operators,
would require a number of iterations that also scales
with $1/\sin(\theta/2)$.
Hence, we get the same scaling as for Grover's algorithm.
\subsection{Connection to Eigenpath Traversal}\label{subsec:ET}

The RBA performs a sequence of reflections to guide the evolution of a quantum state. 
This sequence of reflections can be found by a discretization of an
adiabatic path, where the reflections are applied through the subspaces spanned by the ground states of the instantaneous Hamiltonians chosen. Previous studies have shown that the traversal of an eigenpath
can be accomplished by a scheme resembling the quantum Zeno effect~\cite{boixo2009eigenpath,boixo2010fast,lemieux2021resource}, preparing
the target states with high fidelity by performing a sequence of projective measurements.
Normally, the quantum Zeno effect involves frequently performing projective measurements 
that are all in the same basis, effecting a suppression of the system's own dynamics and \lq\lq{}localizing\rq\rq{} its quantum state in one of the eigensubspaces of the measured observable.
In the context of an adiabatic scheme, however, the measurement basis is continually changed during the computation: in order to drag the system along an adiabatic evolution, the system state is projected onto the ground state of the instantaneous Hamiltonians of the corresponding adiabatic schedule. In the unlikely event of failure, the ground state can be recovered using a rewind procedure~\cite{lemieux2021resource}.

It should be noted that the algorithm by \mbox{Boixo et al.~\cite{boixo2010fast}} also uses reflections along the eigenpath (defined as a sequence of eigenstates $\ket{\psi_s}$, for $0\leq s \leq 1$, of unitary operators $U_s$ or Hamiltonians $H_s$). However, in that scheme, each instantaneous reflection is controlled by an ancilla that is used to flag the desired eigenstate $\ket{G_k}$ in that step. This flag qubit is subsequently measured,  which is, in effect, the implementation of  a binary projective measurement $\{\ketbra{G_k},\mathds{1}-\ketbra{G_k}\}$. In the event of failure, 
another reflection is used to introduce a significant overlap with $\ket{G_k}$.
This is akin to the effect of performing a rewind procedure used by Lemieux et al.~\cite{lemieux2021resource}. These operations are repeated 
until the flag qubit signals the $\ket{G_k}$ has been successfully prepared. Hence, although using reflections, the overall progression 
of the algorithm~\cite{boixo2010fast} results in a quantum Zeno effect--like eigenpath traversal. 
In contrast, the RBA does not rely on intermediate projective measurements. 
Instead, it deploys consecutive reflections around the instantaneous eigenstates 
to traverse the eigenpath without forcing the evolving system state to 
\lq\lq{}localize\rq\rq{} in the instantaneous eigensubspace at each step 
along the schedule.

In what follows, we demonstrate that a reflection around an intermediate 
state gives a better probability of success
than a binary projective measurement defined by the same state.

Let us denote the initial state and the final state by $\ket{I}$ and $\ket{F}$,
respectively. Beginning with $\ket{I}$, we aim to reach $\ket{F}$ with as high a probability as possible. Furthermore, let us introduce a sequence of binary
projective measurements, $\{P_k,\bar{P}_k\}$,
where $P_k:=\ketbra{k}{k}$ and
$\bar{P}_k:=\mathds{1} - P_k$, as well as a sequence of
independent corresponding reflections,
$R_k = \mathds{1}-2\ketbra{k}$, 
for $k \in \{1,...,L\}$.
Using the triangle inequality, given
\begin{align}
  \delta P := |\bra{F}P_k\ket{I}| - |\braket{F}{I} | \geq 0,
\end{align}
we have
\begin{align}
  \delta R := |\bra{F}R_k\ket{I} | - |\braket{F}{I} | \geq 2 \delta P.
\end{align}
In other words,
if an intermediate projective measurement improves the overlap between the initial and the final states, then the corresponding
reflection will cause an even greater overlap.
This also means that we can increase the probability of success
by adding a reflection whenever adding the corresponding projective measurement
is advantageous. Similarly, the use of a telescoping sum implies that
\begin{align}
    | \bra{F}\prod_{k=1}^L R_k \ket{I}| - |\braket{F}{I}| = \sum_{k=1}^L \delta R_k \geq 2 \sum_{k=1}^L \delta P_k,
\end{align}
where 
\begin{align}
    \delta R_k &:= |\bra{F} \prod_{j=1}^k R_j \ket{I}| - |\bra{F}\prod_{j=1}^{k-1}R_j\ket{I} | 
\end{align}    
and
\begin{align}
    \delta P_k &:= | \bra{F} P_k \prod_{j=1}^{k-1} R_j \ket{I} | - | \bra{F}\prod_{j=1}^{k-1}R_j\ket{I} |.
\end{align}
\begin{figure}
    \centering
    \includegraphics[width=0.8\linewidth]{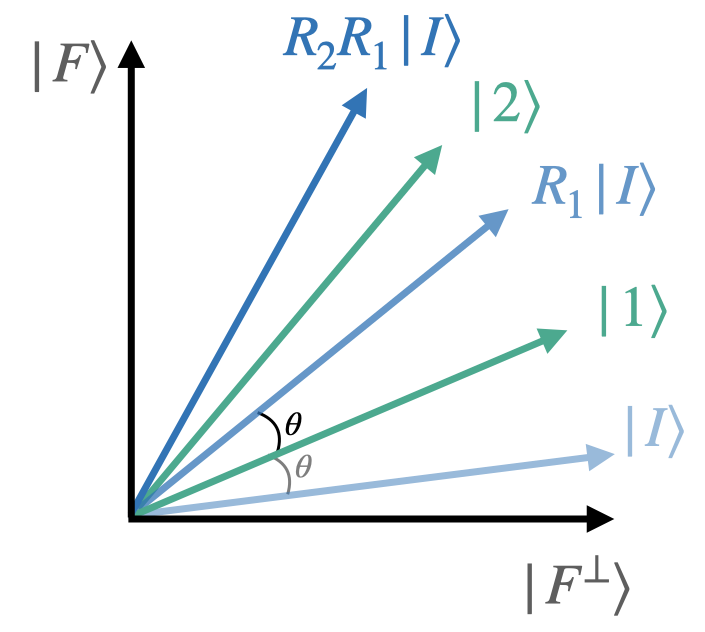}
    \caption{ Visualization of the reflection-based preparation of the target state  when adding reflections increases the overlap with the target state,
   as $|\bra{F}P_2R_1\ket{I} |  > |\bra{F}R_1\ket{I} |  $}
    \label{fig:TwoRefS}
  \end{figure}
  \begin{figure}
    \centering
    \includegraphics[width=0.8\linewidth]{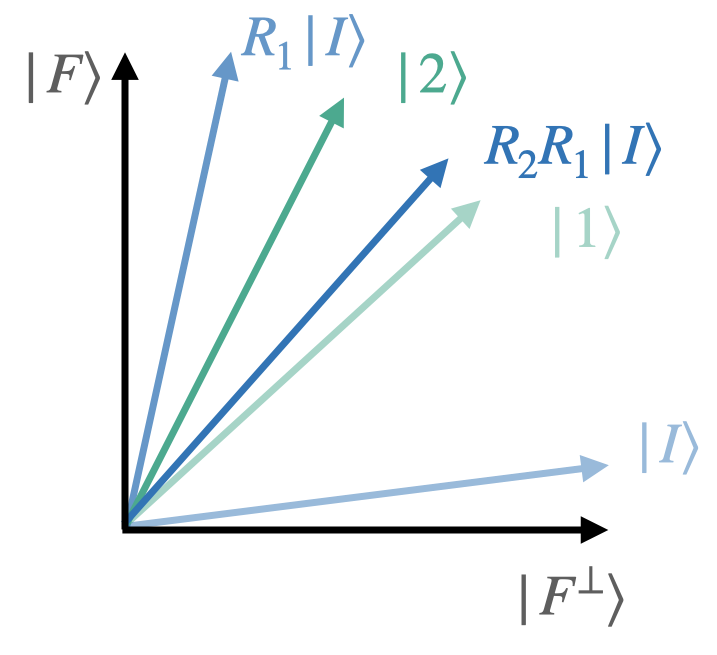}
    \caption{Visualization of the reflection-based preparation of the target state when 
  adding a second reflection decreases the overlap with the target state, as
  $|\bra{F}P_2R_1\ket{I} | < |\bra{F}R_1\ket{I} |$}
    \label{fig:TwoRefF}
  \end{figure}
We illustrate these concepts in Fig.~\ref{fig:TwoRefS} and in Fig.~\ref{fig:TwoRefF}. Figure~\ref{fig:TwoRefS} 
shows when the condition needed to improve the probability of success
is satisfied.
Figure~\ref{fig:TwoRefF} illustrates the case when, even if making two projective
measurements would increase the probability of success as compared to just one, 
the condition for increasing the probability of success by adding a second reflection
is not satisfied.  Note that in the latter case, performing one reflection is actually better than making
two projective measurements to increase the overlap with the target state. 
For details on the assumption that intermediate projective measurements will result in an increase in the overlap,
see the work of Aharonov and   Ta-Shma~\cite{aharonov2007adiabatic} and Boixo et al.~\cite{boixo2009eigenpath}.
  
\section{Performance Analysis of the RBA}

To study the RBA, we compare its performance to Grover's search. 
We use the NP-hard \mbox{\textsc{MAX-2SAT}} problem as the reference problem. 

\subsection{Application to the \textsc{MAX-2SAT} Problem}
In its conjunctive normal form, each \mbox{\textsc{MAX-2SAT}} problem instance 
can be expressed as the Boolean \mbox{formula}
\begin{align}
 \bigwedge_{c\,\in\, \mathcal{C}} \left (\ell_{c_1} \lor \ell_{c_2} \right )\,,
\end{align}
where $\mathcal{C}$ denotes the set of all clauses 
composing the formula. Here, \begin{align}
    \ell_{c_1} , \ell_{c_2} \in \{v_0,\neg v_0 ,..., v _{n-1},\neg v _{n-1}\}
\end{align}
are a pair of distinct literals specifying a clause $c\in \mathcal{C}$, and $v_k\in\{\textsc{True},\textsc{False}\}$, for \mbox{$k=0,1,\dots,n-1$,} are $n$ Boolean variables.  
Notice that $c_1$ and  $c_2$ serve as labels for the two literals making up the clause $c$. For instance, for  the clause $c=\left (\neg v_i \lor v_j\right )$,  $c_1=i$ and  $c_2=j$.
The problem consists in the task of determining the maximum number of clauses  $c\in \mathcal{C} $ that can be simultaneously satisfied by an assignment to the Boolean variables. 

 The \textsc{MAX-2SAT} problem can be recast as the problem of finding the ground state 
 of the $2$-local Hamiltonian 
 \begin{align}
 H_{\text{\tiny $\mathcal{C}$}}&:=\sum_{c\,\in\, 
 \mathcal{C}}
 \left((-1)^{\nu({\ell_{c_1}})}(-1)^{\nu({\ell_{c_2}})}Z_{c_1}Z_{c_2} \right.\nonumber\\
&\quad \quad\quad\left.+ (-1)^{\nu({\ell_{c_1}})} Z_{c_1}+ (-1)^{\nu({\ell_{c_2}})} Z_{c_2}\right),
\end{align}
where $Z_k$ represents the Pauli operator pertaining to variable $v_k$,
\mbox{$\nu({\ell})=0$} if the literal $l$ is not negated, and  $\nu({\ell})=1$ if it is. 
The mapping of the \mbox{\textsc{MAX-2SAT}} problem to the Hamiltonian is such 
that the eigenvalues $+1$ and $-1$  of the 
Pauli-$Z_k$ operator correspond to  the Boolean values \textsc{False}  and \textsc{True}  
of the variable $v_k$, respectively.  
The energy spectrum of this Hamiltonian is such that, every time a clause is not satisfied 
for a given variable assignment, there is a corresponding energy penalty of $+3$, and if a clause is satisfied, its energy contribution is $-1$. Therefore,
the ground states of this Hamiltonian correspond to the variable assignments that satisfy the
maximum number of clauses. By shifting the energy spectrum by $|\mathcal{C}|$, the ground state energy would be equal to zero if the problem is satisfiable, and higher if it is not.  This way, at any iteration, if the outcome of measuring the energy is zero, it is certain that the problem is satisfiable. Also, note that the entire spectrum is upper bounded by $4|\mathcal{C}|$.

In our analysis, the initial state is chosen to be the uniform superposition
\begin{align}\label{eq:allBin}
  \ket{+}^{\otimes n}: = \text{\textsc{Had}}^{\otimes n}\ket{0}^{\otimes n}=\frac{1}{\sqrt{2^{n}}} \sum_{k=0}^{2^{n}-1} \ket{k},
\end{align}
which can be easily generated by applying a Hadamard gate $\text{\textsc{Had}}$ to each qubit.
This state is the ground state of the transverse field Hamiltonian
\begin{align}
  H_{\text{\tiny trans}}:=-\sum_{k=0}^{n-1}X_k,
\end{align}
which is commonly chosen as the initial Hamiltonian of the adiabatic schedule,
where $X_k$ is the Pauli-$X$ operator corresponding to $v_k$. 

 \subsection{Numerical Study}\label{subsec:numStudy}
 Our numerical benchmark includes simulations of the success probability of the RBA and 
a cost analysis in comparison to Grover's search 
in solving small  \mbox{\textsc{MAX-2SAT}} problem instances.
Our results pertain to classically optimized sets of reflection points and the ideal choice of scaling parameters for $H_0$ and $H_1$. Note that, 
in practice, to optimize reflection points along the path, we would need to sample the final energy 
and use a classical optimizer to find a (sub-)optimal schedule for the reflections.

Our numerical simulations use the following steps.

\begin{enumerate}
  \item Perform the ideal normalization as follows:
  \begin{align}
  H_0 &:= 2\pi (H_{\text{\tiny trans}}-E_{\text{\tiny trans}}^0
  \mathds{1})/(E_{\text{\tiny trans}}^{2^n-1}-E_{\text{\tiny trans}}^0), \nonumber\\
  H_1 &:= 2\pi (H_{\text{\tiny $\mathcal{C}$}}-E_{\text{\tiny $\mathcal{C}$}}^0 
  \mathds{1})/(E_{\text{\tiny $\mathcal{C}$}}^{2^n-1}-E_{\text{\tiny $\mathcal{C}$}}^0) ,
  \end{align}
  where $E_{\text{\tiny trans}}^j$ and $E_{\text{\tiny $\mathcal{C}$}}^{j}$ for 
  $j=0,\dots,(2^n-1)$ are the eigenvalues of energies of the Hamiltonian 
   $H_{\text{\tiny trans}}$ and $H_{\text{\tiny $\mathcal{C}$}}$, respectively. 
  \item Set the total number of reflection points \mbox{$L=1$.} 
  \item Use the Nelder--Mead optimization protocol to find a set $\{w_1,...,w_L\}$ of reflection points (by minimizing the probability of failure), assuming energy thresholds between $E_{w_k}^0$ and $E_{w_k}^1$ for all $k=1,...,L$,  and compute the time to solution (TTS). 
  \item Repeat the previous step by performing \mbox{$L=2,3,4,...$} reflections until a minimum TTS is observed.  
\end{enumerate}

In practice, we do not have access to $E_{w_k}^0$ and $E_{w_k}^1$ in step 3. However, we may use the known values of both $E_0^0$ and $E_0^1$, and an upper bound for $E_1^0$ determined using other techniques (e.g., polynomial-time approximation schemes~\cite{lewin2002improved} or local heuristic search methods~\cite{li2009theory}), to construct a concave threshold function with respect to $w$ and tune the concavity of the function to estimate tight upper bounds for $E_{w_k}^0$ for all reflection points $w_k$. We also note that the normalization in step 1 is considered ``ideal'' since we do not have access to $E_1^0$ or $E_1^{2^n-1}$. However, we can use $4|\mathcal{C}|$, or another heuristically found bound, as an upper bound for $\left(E_1^{2^n-1}-E_1^0\right)$.

For small problem instances with the number of variables \mbox{$n \in \{5,...,13\}$}, we 
compute the probability of success $p_s$ (i.e, the modulus squared of the overlap between 
the final computational state and the target state), given the
above choices of threshold, reflection points, and normalization. The TTS is then given by
\begin{align}\label{eq:TTS-RBA}
  \text{TTS}_{\tiny \text{RBA}}& = \frac{\log\epsilon}{\log(1-p_s)}\sum_{k=0}^L \frac{2r}{\Delta_k}\,,
\end{align}
where $\Delta_k = E^1_{w_k}-E^0_{w_k} $ is the normalized instantaneous energy gap, $r:=|\mathcal{C}|/n$, and 
$\epsilon$ is the overall allowable probability of failure. In our simulations, we set $\epsilon=0.1$.

In Eq.~(\ref{eq:TTS-RBA}), the cost of each trial of the algorithm is 
$\sum_{k=0}^L \frac{2r}{\Delta_k}$, as QPE requires a number of queries to Hamiltonian simulation unitaries
that scales with the inverse of the precision needed 
to distinguish between the energies of the first excited state and the ground state. 
Moreover, each unitary has a gate depth that scales with the
\mbox{ratio $r$.} 

The TTS for Grover's algorithm is given by
\begin{align}
  \text{TTS}_{\tiny \text{Grover}}& = \frac{\log\epsilon}{\log(1-p_s)}\sum_{k=1}^{n_{it}} \frac{2r}{\Delta_{T}}
  \nonumber\\& = \frac{\log\epsilon}{\log(1-p_s)} \frac{2n_{it} r}{\Delta_{T}}, 
\end{align}
where $\Delta_{T} = 2\pi (E^1_{\text{\tiny $\mathcal{C}$}}-E_{\text{\tiny $\mathcal{C}$}}^0)/(E_{\text{\tiny $\mathcal{C}$}}^{2^n-1}-E_{\text{\tiny $\mathcal{C}$}}^0)$
is the normalized gap pertaining to the target Hamiltonian. 
Note that we ignore the additional cost \mbox{of $R_1^G$}.

\begin{figure}
    \centering
    \includegraphics[width=\linewidth]{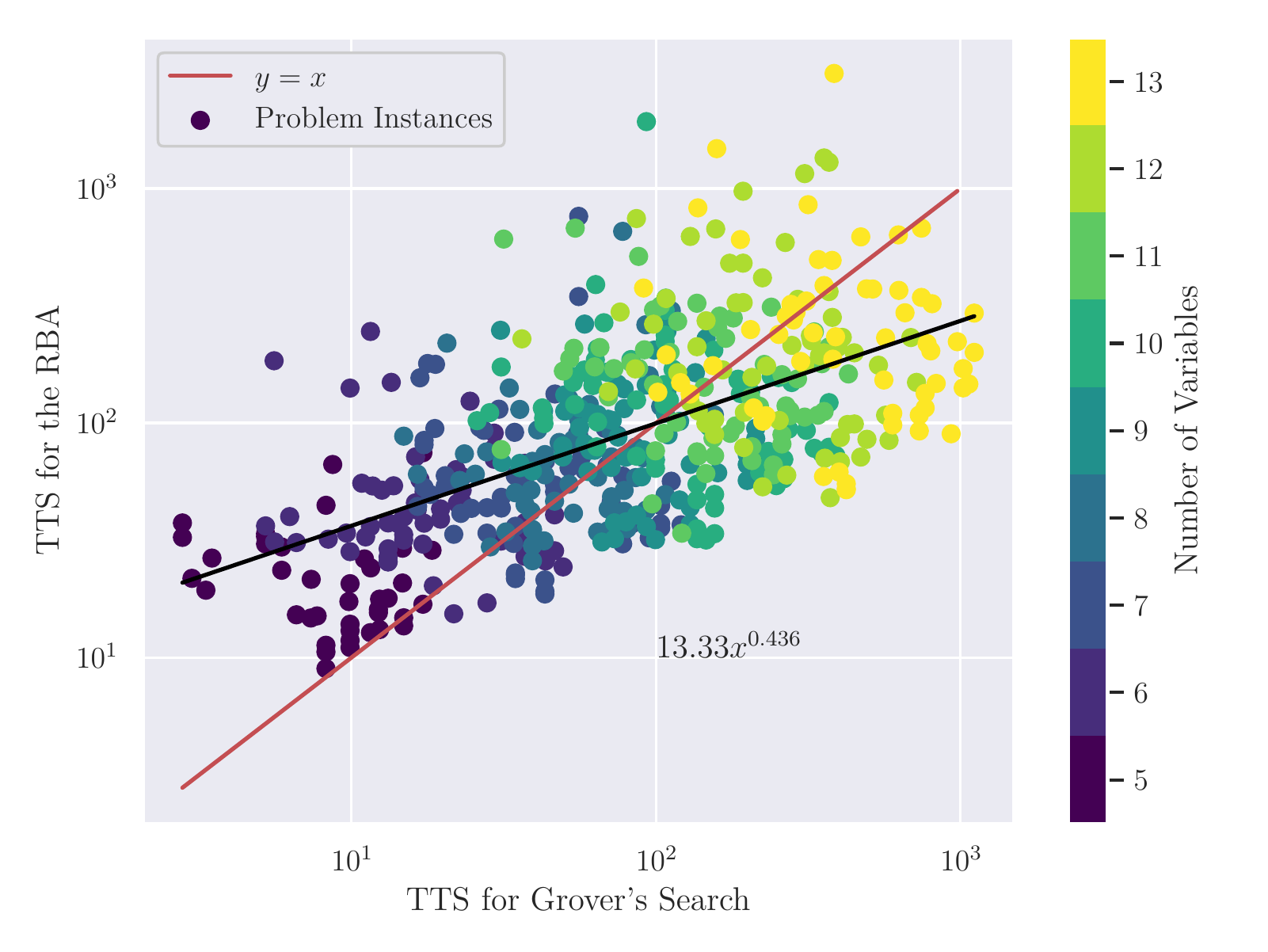}
    \caption{Improvement in TTS for the RBA compared to Grover's search. The colours indicate the number of variables used for each problem instance.}
    \label{fig:TTSnumVar}
  \end{figure}

\begin{figure}
    \centering
    \includegraphics[width=\linewidth]{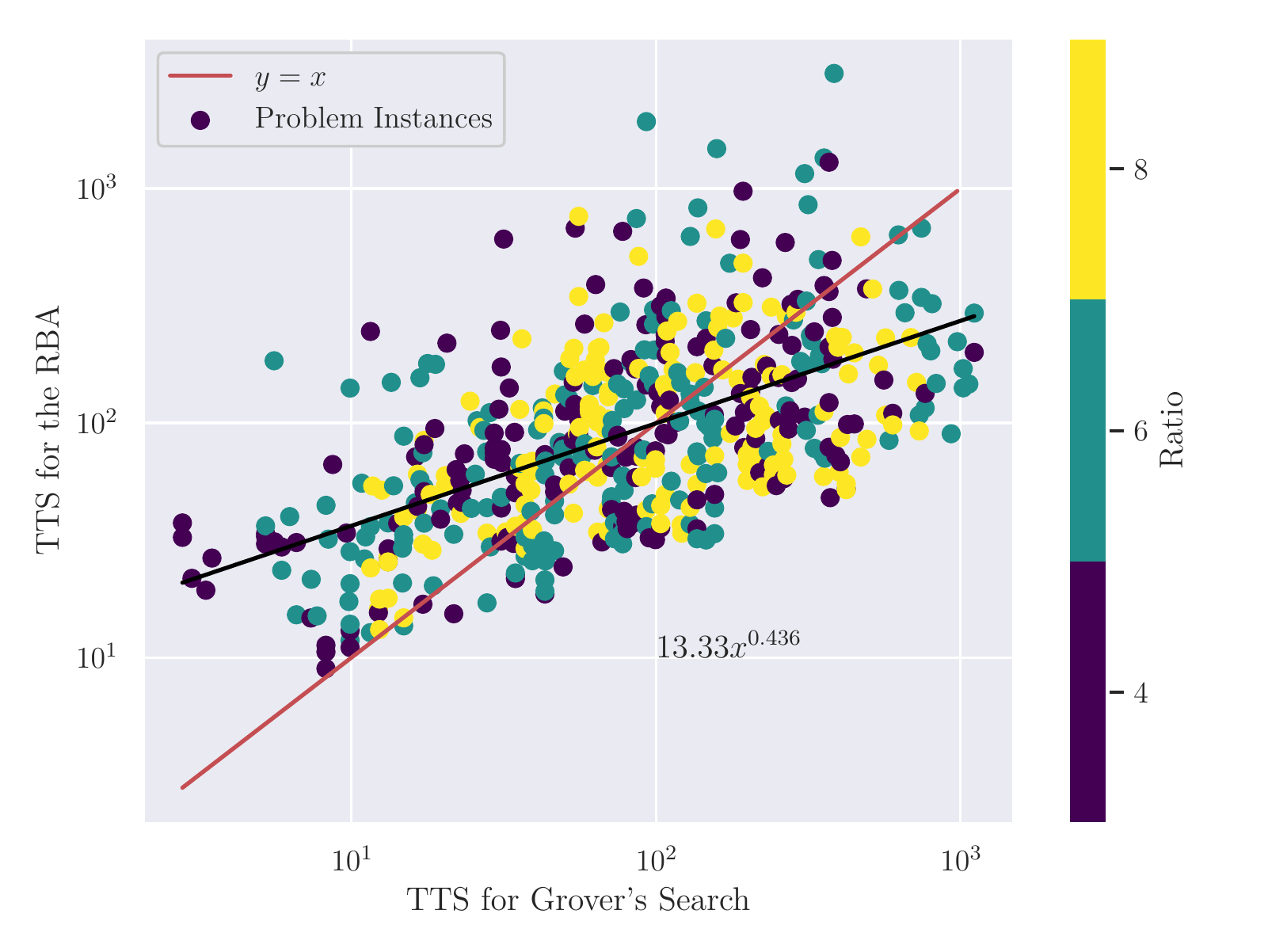}
    \caption{Improvement in TTS for the RBA compared to Grover's search. The colours indicate the ratio used for each problem instance.}
    \label{fig:TTSratio}
  \end{figure}

We generate $20$ distinct random instances for each problem size and each $r\in \{4,6,8\}$ (with the exception of the combination $n=5$ and $r=8$, 
in which case there is only one possible instance), and compute the corresponding 
TTS values. 
We plot the results we obtain for the RBA against those we obtain for Grover's search in
Fig.~\ref{fig:TTSnumVar} and in Fig.~\ref{fig:TTSratio}, where we can see a trend indicating a speed-up relative to Grover's search. The TTS depends on both $n$ and $r$. The impact of both of these parameters 
is shown in colour, with the colour bars indicating the number of variables in
Fig.~\ref{fig:TTSnumVar} and the ratio in Fig.~\ref{fig:TTSratio}.
  
\begin{figure}
  \centering
  \includegraphics[width=\linewidth]{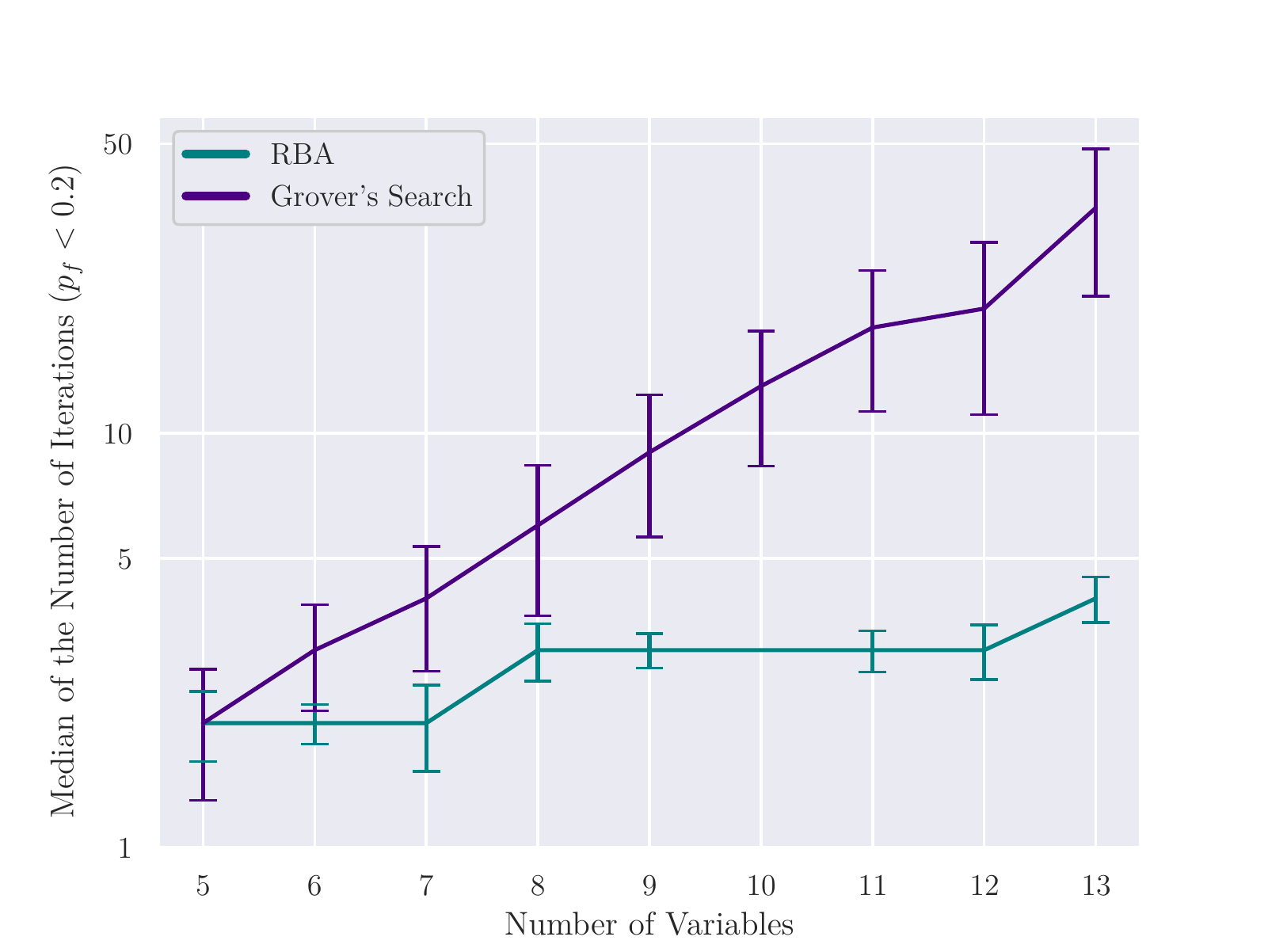}
  \caption{Median of the number of iterations versus the number variables for the RBA and Grover's search that is needed to obtain a probability of failure below a value of 0.2}
  \label{fig:failProb}
\end{figure}

These figures illustrate that, for small problem instances, 
Grover's algorithm performs better, but, as the
problem becomes harder, the RBA tends to outperform
Grover's algorithm. The reason is that the number of iterations required for
our algorithm to reach a high probability of success grows more slowly
with the system size than in Grover's search, as shown in  Fig.~\ref{fig:failProb},
but each reflection has a higher
cost (due to there being a smaller gap for the instantaneous Hamiltonian 
in comparison to $H_0$ and $H_1$).

Figure~\ref{fig:failProb} shows the median values of the number
of iterations required for both algorithms to reach a target probability of failure
below $0.2$. As expected, for Grover's algorithm, we observe an exponential dependence of the number of iterations
on the number of variables. For the RBA, the nature of this relationship is not readily apparent. As the number of iterations is discrete and we only include simulations for 
problem instances with up to 13 variables, it is hard to extract a clear trend. However,
we observe a significant decrease in the number of iterations as compared to Grover's search.

The TTS scaling of the RBA could potentially be reduced 
by using the method of 
\mbox{qubitization~\cite{berry2018improved,lemieux2021resource,low2019hamiltonian,poulin2018quantum}.} 
Indeed, qubitization would change the overlap between successive states as well as the gap from $\Delta$ to
$\arccos(1-\frac{\Delta}{2\pi})$, which has a greater impact for smaller gap values~\cite{lemieux2021resource}. This could potentially result in  a further 
improvement of the RBA over Grover's search.

For a hybrid approach, in which we optimize
the choice of each $w_k$ using a
classical optimizer, we study the phenomenon of the barren plateau. To do so, we compute the variance
$\text{var}\left(\frac{\partial E}{\partial w_i} \right)$
when $L=2$. We consider the same problem instances as earlier. 
To sample the variation in energy based on the positions of
the reflections, we take 5000 random points uniformly distributed from within 
the intervals $\left(\frac{1}{6},\frac{1}{2}\right)$ and $\left(\frac{1}{2},\frac{5}{6}\right)$.
\begin{figure}
  \centering
  \includegraphics[width=\linewidth]{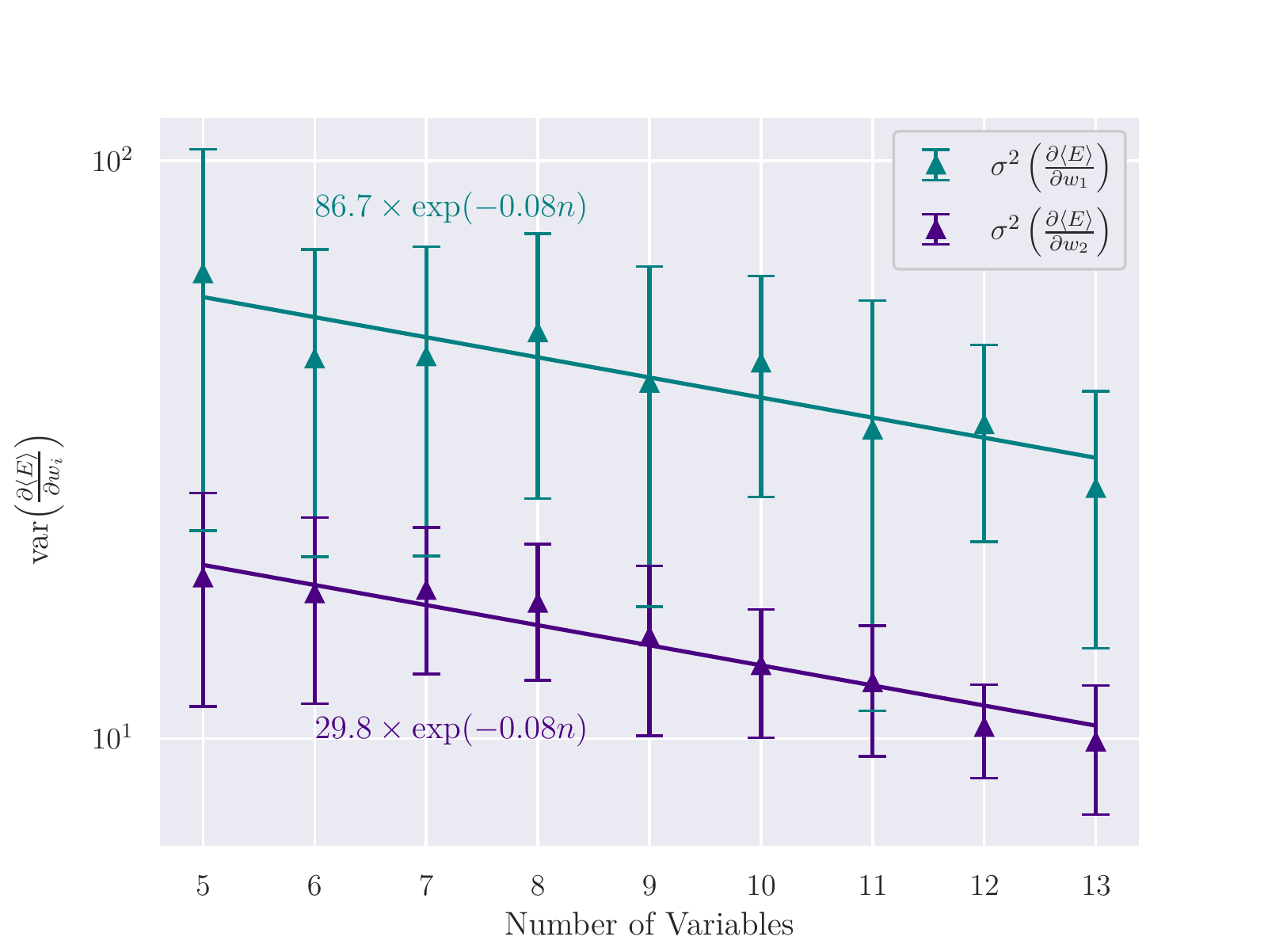}
  \caption{Variance of the partial derivative of the expected energy versus the number of variables for the RBA, demonstrating an exponential decay associated with the existence of a barren plateau for a schedule with two reflections}
  \label{fig:BPdw1}
\end{figure}
In Fig.~\ref{fig:BPdw1}, we observe an exponential decay with a rate of $0.08$ for the median values of gradient variances. This is in contrast to the rate of $0.685$ obtained for the random parameterized quantum circuits studied by McClean et al.~\cite{mcclean2018barren}. The error bars indicate the standard deviation.

\begin{figure}
    \centering
    \includegraphics[width=\linewidth]{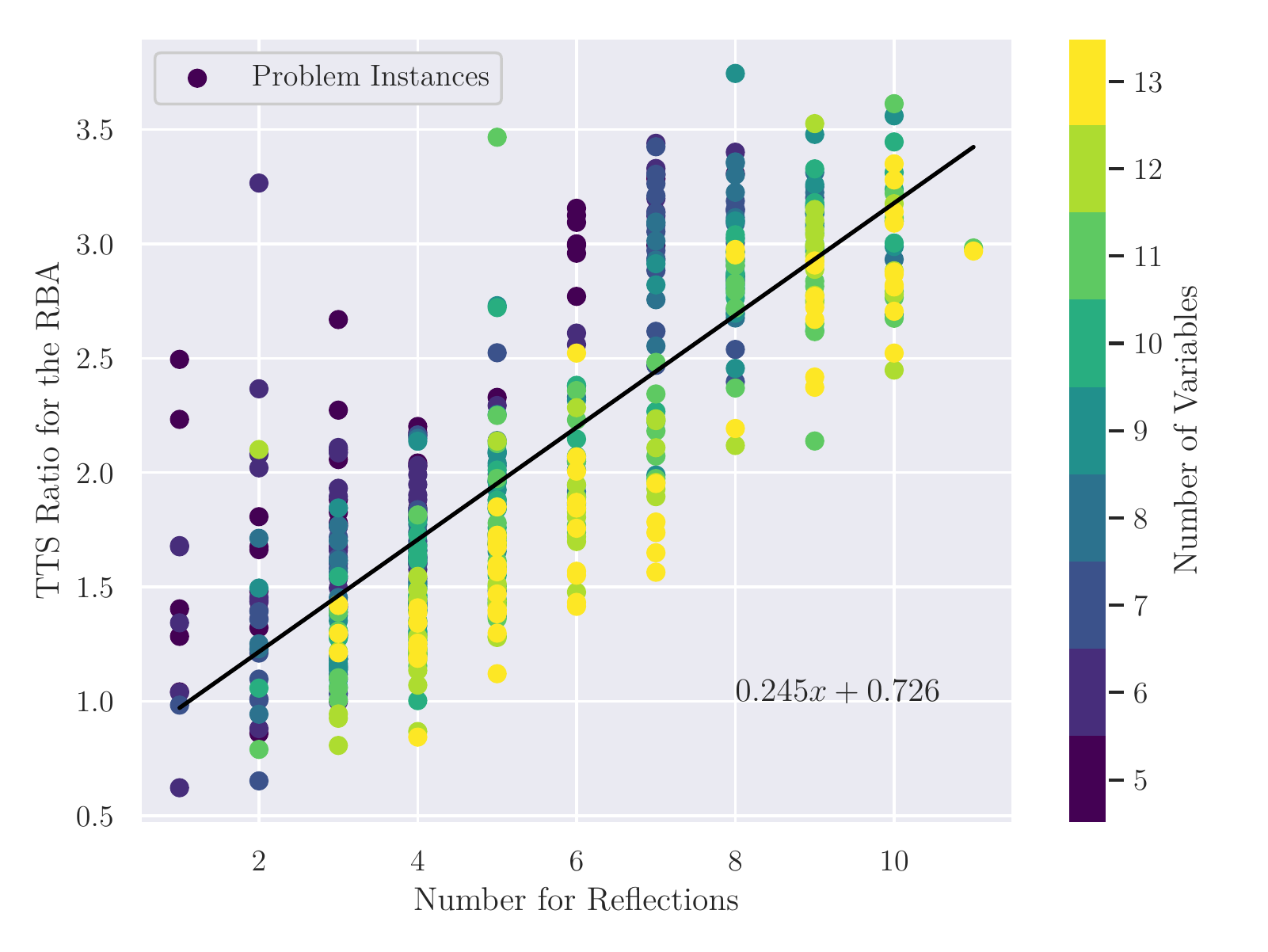}
    \caption{Time-to-solution ratio of the non-optimized RBA over one that is  optimized versus the number of reflections. The non-optimized RBA uses a schedule with equidistant reflection points. The optimized RBA uses a schedule that minimizes the probability of failure.
  A decrease in the TTS resulting from optimizing the positions of the reflections can be seen. The colours indicate the number of variables used for each problem instance.}
    \label{fig:RTTSop}
\end{figure}

\begin{figure}
    \centering
    \includegraphics[width=\linewidth]{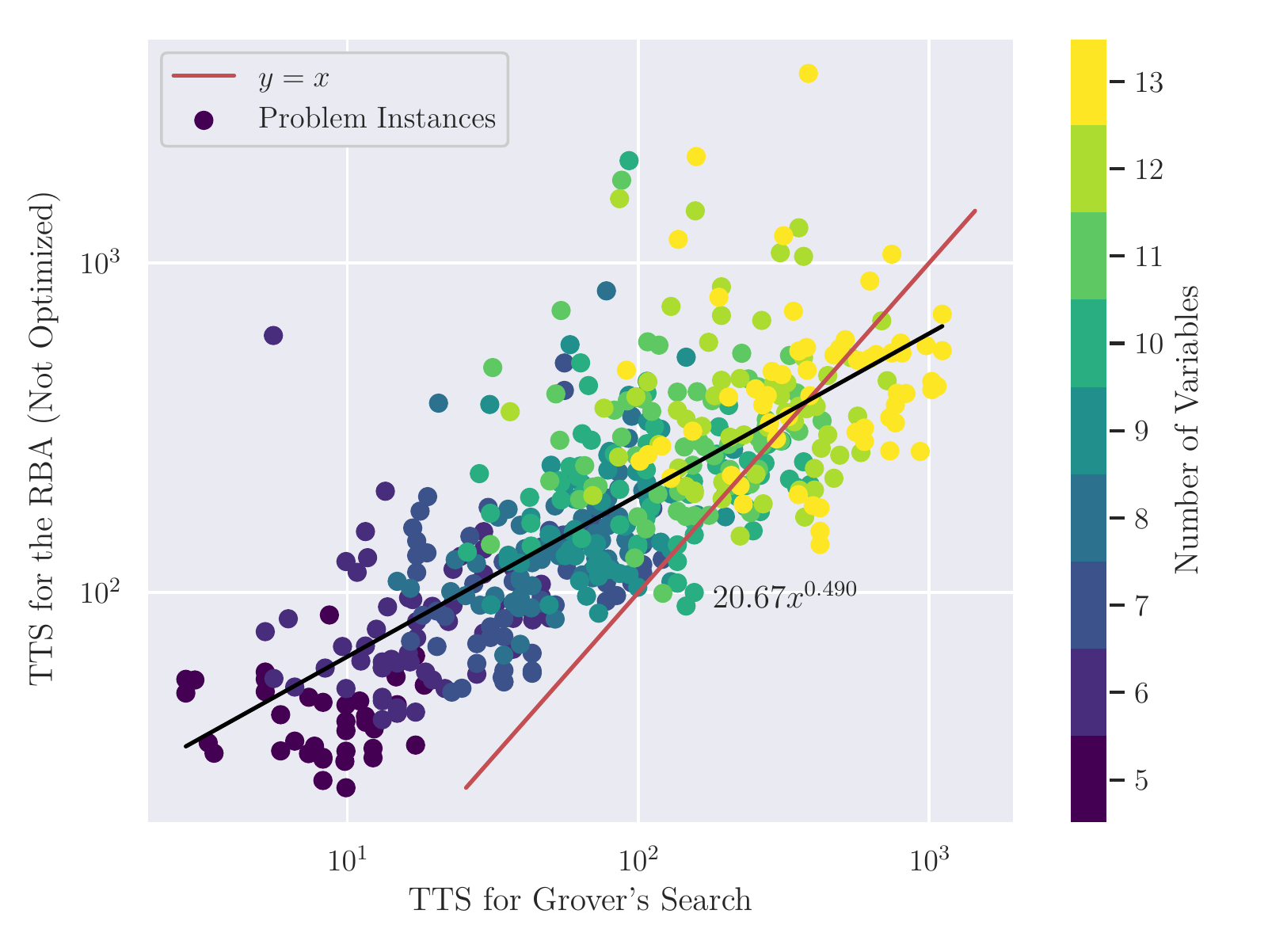}
    \caption{Time to solution for the non-optimized RBA versus the TTS for Grover's search.
  Optimization of the reflection points is not performed; the figure is otherwise equivalent to Fig.~\ref{fig:TTSnumVar}.
  The non-optimized RBA uses a schedule with equidistant reflection points. The colours indicate the number of variables used for each problem instance.}
    \label{fig:TTSratioOP}
\end{figure}

Figure~\ref{fig:RTTSop}
shows the TTS ratio of the non-optimized RBA (i.e., using equidistant reflection points between $0$ and $1$) over one that optimizes the reflection points versus the number of reflections (i.e., the advantage offered by the classical optimizer).
 We observe that not performing the optimization adds, on average, a linear factor
with the number of reflections. Notice that, in some cases,
the non-optimized version has a smaller TTS. This is because we minimize the probability of failure and not the TTS. Indeed, for some cases, the smaller probabilities of success exist alongside 
larger gaps, which can result in a smaller TTS. Figure~\ref{fig:TTSratioOP} shows the TTS for the non-optimized RBA compared to the TTS for Grover's search. We observe that, even with a schedule with equidistant reflection points, the RBA offers a speed-up over Grover's search. Figure~\ref{fig:TTSratioOP} is equivalent to Fig.~\ref{fig:TTSnumVar}, except that it does not include the optimization of the reflection points. 

Finally, we empirically study
the impact of having an energy threshold between $E^1_{w_k}$ and $E^2_{w_k}$ and observe that
it does not result in a significant increase in the TTS. In fact, in some cases, it
results in a decrease in the TTS, especially for instances 
that exhibit a closing energy gap toward the end of the adiabatic path.
This is due to there being
a reduced-precision requirement needed to implement QPE. Indeed, in this case, the precision requirement scales inversely with $(E^2_{w_k}-E^0_{w_k})$ instead of with $(E^1_{w_k}-E^0_{w_k})$.
Thus, we expect that the RBA is not significantly sensitive to optimal choices of energy thresholds.

\section{Conclusion}

We have presented a new circuit-model quantum algorithm for preparing the ground state of a target problem Hamiltonian. The reflection-based adiabatic algorithm (RBA) is based on a combination of concepts from  Grover's search and adiabatic quantum computation. We have shown that the resource requirements for the RBA to reach the target state with a high probability of success scale favourably compared to Grover's search in solving MAX-$2$SAT problem instances. We used the time-to-solution metric (TTS), that is, the time needed to find a solution for a particular problem instance with high confidence (e.g., a rate of 90\%). Although our algorithm has a higher implementation cost per reflection, as the probability of success increases faster with the number of reflections than it does in Grover's search, the TTS scales better. 

The complexity analysis made by Boixo et al.~\cite{boixo2010fast} suggests that path traversal algorithms could outperform Grover's algorithm in solving search and optimization problems. In the present work, we have provided numerical evidence supporting this claim. Although the RBA does not employ projective measurements but instead uses reflections to traverse the eigenpath, the inequalities from Sec.~\ref{subsec:ET} suggest that its query complexity is not worse than given by Boixo et al.~\cite{boixo2010fast}.

Previous work on discrete adiabatic state preparation using continual instantaneous projective measurements~\cite{lemieux2021resource} incorporated techniques such as qubitization~\cite{low2019hamiltonian} and a rewind procedure.
We expect that qubitization would also reduce the implementation cost of the reflections in the RBA. However, a rewind procedure is not helpful in a unitary scheme (i.e., without projective measurements); thus, we have not employed such a procedure in our implementation, which benefits from full quantum parallelism.
Indeed, it is not only the ground states along an adiabatic path that contribute to the probability amplitude of obtaining the desired target state, but so do the excited states. The contribution of the excited states becomes especially important in schemes where the gap closes toward the end of the adiabatic schedule, which is the case for degenerate instances of \textsc{MAX}-$k$\textsc{SAT} problems. 
Moreover, in real-world experiments, the measurements required in measurement-based adiabatic schemes are generally slower and much more prone to suffering from errors than are unitary quantum gates. In light of our findings, we also believe that the heuristic use of the walk operator
by Lemieux et al.~\cite{lemieux2020efficient} works well without employing phase randomization~\cite{boixo2009eigenpath}, because the walk operator consists of reflections. 

In looking toward potential implementations of a hybrid RBA scheme supplemented by a classical optimization loop, we have investigated the barren plateau phenomenon. 
Our simulations indicate that there is a small exponential decay associated with the existence of a barren plateau, with a rate of $0.08$.
However, the decrease in the TTS resulting from optimizing the positions of reflections along an adiabatic path scales linearly with the number of reflections. 
Since the number of reflections employed in the RBA scales more slowly than that 
in Grover's search, we expect that, even without optimization, the RBA algorithm will provide a greater speed-up, on average, than  Grover's search.

\section*{Acknowledgement}
The authors thank our editor, Marko Bucyk, for his careful review and editing this manuscript, 
Simon Verret and Gili Rosenberg for helpful discussions. P.~R.~additionally acknowledges the financial support of Mike and Ophelia Lazaridis, and Innovation, Science and Economic Development Canada.
\newpage
\bibliographystyle{plainurl}
\bibliography{ref.bib}
\end{document}